\documentclass[aps,prd,11pt,showpacs,preprintnumbers,nofootinbib,superscriptaddress]{revtex4}
\def\nbox#1#2{\vcenter{\hrule \hbox{\vrule height#2in
\kern#1in \vrule} \hrule}}
\def\sq{\,\raise.5pt\hbox{$\nbox{.09}{.09}$}\,}
\def\sqb{\,\raise.5pt\hbox{$\overline{\nbox{.09}{.09}}$}\,}

\newcommand{\bea}{\begin{eqnarray}}
\newcommand{\eea}{\end{eqnarray}}
\newcommand{\be}{\begin{equation}}
\newcommand{\ee}{\end{equation}}
\newcommand{\bes}{\begin{subequations}}
\newcommand{\ees}{\end{subequations}}
\def\la{\langle}
\def\ra{\rangle}

\usepackage{amssymb,amsmath}
\usepackage{graphicx}

\begin{document}

\title{Breakdown of the semiclassical approximation during the early stages of preheating}
\author{Paul R. Anderson}
\email{anderson@wfu.edu}
\affiliation{Department of Physics, Wake Forest University, Winston-Salem, North Carolina, 27109, USA}
\author{Carmen Molina-Par\'{\i}s}
\email{carmen@maths.leeds.ac.uk}
\affiliation{Department of Applied Mathematics,
University of Leeds, Leeds LS2 9JT, UK}
\author{Dillon H. Sanders}
\affiliation{Department of Physics,
Wake Forest University, Winston-Salem, North Carolina, 27109, USA}
\affiliation{Department of Nuclear Engineering, North Carolina State University,
Raleigh, NC 27695, USA}
\email{dhsander@ncsu.edu}

\begin{abstract}
\vskip .3cm

The validity of the semiclassical approximation is investigated during the preheating phase in models of
chaotic inflation using a modification of a criterion previously proposed for semiclassical gravity.
If the modified criterion is violated then fluctuations of the two-point function for the quantum fields are
large and the semiclassical approximation is not valid.  Evidence is provided that the semiclassical
approximation breaks down during the early stages of preheating, well before either scattering effects or backreaction effects
are important.

\end{abstract}


\maketitle

\vfill\eject

The semiclassical approximation has been used to study the effects of quantized fields on a classical background field in a wide variety of scenarios including black hole evaporation~\cite{hawking}, the decay of an electric field due to the Schwinger effect~\cite{kluger}, heavy ion collisions in nuclear physics~\cite{kluger-2}, and preheating in chaotic inflation~\cite{kls1,amec}.  It is expected to be valid in cases where quantum effects are small, such as the initial stages of the evaporation of a solar mass black hole.
At the opposite end of the spectrum is the case of preheating in models of chaotic inflation.  Preheating occurs immediately after the inflationary phase and is a period in which the rate of particle production is extremely rapid, resulting in strong backreaction effects upon the inflaton field~\cite{kls2,d-k}.
It is not known whether the predictions of the semiclassical approximation can be trusted when quantum effects are so large.

The semiclassical backreaction equations for quantum fields coupled to a classical background field arise out of the one loop effective action
for that field~\cite{b-d-book}.  As such they would typically be expected to break
down when backreaction effects are large and terms coming from higher loops may be important.  One way around this is to use a large $N$ expansion where $N$ is the number
of identical quantum fields.  The semiclassical backreaction equations become exact in the limit $N \rightarrow \infty$.  This expansion has been used in cases such as preheating~\cite{amec,boy} where backreaction effects are significant.

For the semiclassical approximation to be valid, quantum fluctuations about the mean of whatever quantity couples the quantum fields to the
classical background field(s) must be small.
One way to characterize these fluctuations in semiclassical gravity is through the two-point function for the energy-momentum tensor.  However,
 for the symmetric part of this two-point function there can be state-dependent divergences~\cite{wu-ford}, and different renormalization
 schemes can yield different results when the points come together~\cite{phillips-hu}.
  To overcome these difficulties a criterion was given in~\cite{amm-validity} that
relates the validity of the semiclassical approximation in gravity to the stability of solutions to the linear response equation, which results when the semiclassical backreaction
equation is perturbed about a solution to that equation.  The linear response equation has a term which involves the perturbed energy-momentum tensor, so renormalization proceeds in the usual way and there are no state dependent divergences.

The criterion states that the large $N$ semiclassical approximation in gravity will break down if any linearized gauge invariant quantity constructed from
 solutions to the linear response equations with finite non-singular initial data, grows without bound.
It has been shown to be satisfied  for massive and massless free scalar fields in flat space in the Minkowski vacuum state~\cite{amm-validity}, and for conformally
invariant free fields in the expanding part of de Sitter space, when spatially
 flat coordinates are used, the fields are in the Bunch-Davies state, and scalar perturbations are considered~\cite{amm-desitter}.
Tensor perturbations for conformally invariant free fields were investigated in~\cite{ford-et-al} and
it was found that they are bounded, so the criterion is satisfied in that case as well.

In this paper we continue an investigation begun in~\cite{mg-procedings},
where we adapted the criterion in~\cite{amm-validity} to check the validity of the semiclassical approximation in models of preheating in chaotic inflation,
in which rapid damping of the inflaton field occurs, and found evidence that quantum fluctuations are large in between the two periods of rapid damping.
Here we study in great detail the relationship between
solutions to the linear response equation and quantum fluctuations, and use the results to relate the size of the fluctuations to the particle production rate.
We also include a case in which there is no rapid damping.  We find evidence that quantum fluctuations are large and the semiclassical approximation breaks
down whenever the particle production rate is high, including during the early stages of preheating when scattering effects ignored in our model and backreaction effects on the inflaton field are small.

We consider a model of chaotic inflation for which the inflaton field $\phi$ is coupled to $N$ identical massless scalar fields $\psi_i$ with a coupling of the form
$ \sum_{i=1}^N g^2 \phi^2 \psi_i^2 $.
Full backreaction effects for this coupling have been investigated in
detail in Refs.~\cite{kls1,amec,random} (although not all of
these were in the context of the large $N$ expansion or for massless quantum fields).
After a standard rescaling
of the coupling constant $g$~\cite{amec}, the problem reduces to the coupling of the inflaton field to a single scalar field $\psi$, and the semiclassical backreaction equation for
the inflaton field is $\Box \phi - (m^2 + g^2 \langle \psi^2 \rangle) \phi = 0$.
As in~\cite{amec} we work in a flat space background and consider only homogeneous and isotropic solutions for $\phi$.

The mass of the inflaton field can be scaled out of the equations using $ t \rightarrow mt$  and $\phi \rightarrow \phi/m $,
with similar changes of variable for other relevant quantities.  (See~\cite{amec} for details.)  The result is
\bes \bea
& & \ddot{\phi} + (1 + g^2 \langle \psi^2 \rangle) \phi = 0 \;, \label{sbe} \\
& & \langle \psi^2 \rangle = \frac{1}{2 \pi^2} \int_0^\epsilon dk k^2 \left(|f_k(t)|^2 - \frac{1}{2 k} \right) \nonumber \\
& &  + \frac{1}{2 \pi^2} \int_\epsilon^\infty dk k^2 \left(|f_k(t)|^2 - \frac{1}{2 k} + \frac{g^2 \phi^2}{4 k^3} \right) \nonumber \\
& &  - \frac{g^2 \phi^2}{8 \pi^2} \left[ 1 - \log \left(\frac{2 \epsilon}{M} \right) \right] \;,  \label{psi2-def}  \\
& &  \ddot{f}_k + (k^2 + g^2 \phi^2) f_k = 0   \;.  \label{mode-eq}
\eea \label{se-eqs} \ees
Note that $\la \psi^2 \ra$ is independent of the positive constant $\epsilon$ and that $M$ is a mass scale which typically enters when computing
renormalized quantities for massless fields~\cite{amec}.

The linear response equation can be derived as in~\cite{amm-validity} by taking a
second variation of the effective action.  The result is
\bes \bea && (\Box - m^2 - g^2
\langle \psi^2 \rangle) \delta \phi - g^2 \delta \langle \psi^2
\rangle \phi =0  \; ,
\label{general-linear-response}  \\
&& \delta \langle \psi^2 \rangle = - i g^2 \int d^4 x'
\; \phi(x') \delta \phi(x') \theta(t-t') 
\, \langle [ \psi^2(x), \psi^2(x')] \rangle  + \delta \la \psi^2 \ra_{\rm SD}
\; .
\label{delta-psi-si} \eea \ees
Here $\delta \la \psi^2 \ra_{\rm SD}$ comes from a variation in
the state of the quantum field.

The linear response equation can also be derived by
perturbing the semiclassical backreaction equation about one of its solutions.  We illustrate this
for homogeneous and isotropic perturbations.
The equation for the
inflaton field~\eqref{sbe} and the mode equation~\eqref{mode-eq} are perturbed in the usual way, keeping
quantities that are first order in $\delta
\phi$ and $\delta f_k$.  The perturbed mode equation is then solved in terms
of the solutions to~\eqref{se-eqs}
with the result:
\bea
&& \delta f_k = A_k f_k + B_k f_k^*  + 2 g^2 i \int_0^t dt'\; \phi(t') \delta  \phi(t') f_k(t')
\, [f_k^*(t) f_k(t')-f_k(t) f_k^*(t')] \;. \label{f-solution}
\eea
The coefficients $A_k$ and $B_k$ are related to a change of state and are fixed by the initial values of
$\delta f_k$ and its first derivative.  Such a change in
state (as pointed out in~\cite{prv} for semiclassical gravity) must occur if the original state is a second order or higher
adiabatic state~\cite{b-d-book}.

The linear response equation in this case is
\bes \bea
& & \delta \ddot{\phi} + (1 + g^2 \langle \psi^2 \rangle) \delta \phi + g^2 \phi \delta \langle \psi^2 \rangle = 0 \;, \label{lin-resp-spec} \\
& & \delta \langle \psi^2 \rangle = \frac{1}{2 \pi^2} \int_0^\epsilon dk k^2 \left(f_k \delta f_k^*  + f_k^* \delta f_k \right)  \nonumber \\
& & + \frac{1}{2 \pi^2} \int_\epsilon^\infty dk k^2 \left(f_k  \delta f_k^*  + f_k^* \delta f_k + \frac{ g^2 \phi \, \delta \phi}{2 k^3} \right)
   \nonumber \\
   & &  - \frac{g^2 \phi \, \delta \phi}{4 \pi^2} \left[ 1 - \log \left(\frac{2 \epsilon}{M} \right) \right] \;. \label{delta-psi2} \eea \ees

For the fourth order adiabatic states used in~\cite{amec}, we find that $ A_k = 0 $ to linear order.  An
explicit expression for $ B_k$ can easily be obtained but we will not
display it here.

If one can find solutions to~\eqref{se-eqs} then it is easy to generate
approximate solutions to~\eqref{general-linear-response}.  One
simply takes two solutions, $\phi_1$ and $\phi_2$, which have nearly
the same values at the initial time $t = 0$, and evolves them
numerically in time. If we define the difference between the solutions
to be $\delta \phi_e \equiv \phi_2 - \phi_1$, then $\delta \phi_e$
satisfies the equation:
\be \delta \ddot{\phi}_e + (1 + g^2
\langle \psi^2 \rangle_1) \delta \phi_e + g^2 (\langle \psi^2 \rangle_2
- \langle \psi^2 \rangle_1) (\phi_1 + \delta \phi_e) \;
. \label{delta-phi-exact} \ee
The linear response equation~\eqref{lin-resp-spec} in
this case is
\be \delta \ddot{\phi} + (1 + g^2 \langle \psi^2 \rangle_1) \delta \phi + g^2
\delta \la \psi^2 \ra|_{\phi = \phi_1} \, \phi_1 = 0
\; . \label{specific-linear-response} \ee
Note that the first two terms in these equations have the same form.
  Thus, $\delta \phi_e$, which is a solution to~\eqref{delta-phi-exact}, is also an
approximate solution to~\eqref{specific-linear-response} so long
as the amplitude of the oscillations of $\phi_1$ is much larger than the amplitude
of oscillations of $\delta \phi_e$ and $\delta \la \psi^2 \ra|_{\stackrel{\phi = \phi_1}{\delta \phi = \delta \phi_e}} \approx \langle \psi^2 \rangle_2 - \langle \psi^2
\rangle_1 $.

 Because of the structure of Eq.~\eqref{specific-linear-response} it is possible to go further and separate out the part of the perturbation driven by $\delta \la \psi^2 \ra$.
 For simplicity choose the starting values for $\phi_2$ and $\phi_1$ such that $\phi_2(0) = \phi_{20}$, $\phi_1(0) = \phi_{10}$, and $\dot{\phi}_2(0) = \dot{\phi}_1(0) = 0$.
 Then let
 \be \delta \phi_e = \phi_2 - \phi_1 = c \phi_1 + \delta \phi_c  \;, \label{phic-def} \ee
 with $ c = (\phi_{20} - \phi_{10})/\phi_{10}$.
 Substituting into~\eqref{specific-linear-response} and using~\eqref{sbe}  one finds that if $\delta \phi_e$ is an approximate
  solution to~\eqref{specific-linear-response}, then $\delta \phi_c$ is an approximate solution to the equation
 \bea 
 & & \delta \ddot{\phi}_c + (1+ g^2 \la \psi^2 \ra_1 ) \delta \phi_c + g^2 \phi_1 \delta \la \psi^2 \ra\vert_{\stackrel{\phi=\phi_1}{\delta \phi = \delta \phi_c}} 
  = - g^2 \phi_1 \delta \la \psi^2 \ra\vert_{\stackrel{\phi = \phi_1}{\delta \phi = c \phi_1}}  \;. \label{phic-eq}
   \eea
Thus the equation for $\delta \phi_c$ is the same as that for $\delta \phi$ except that there is a source term
 which depends on $\delta \la \psi^2 \ra$ in~\eqref{delta-psi2} evaluated with $\phi=\phi_1$ and $\delta \phi = c \phi_1$.  
 Since the initial conditions are $\delta \phi_c(0) = \delta \dot{\phi}_c(0) = 0$, at early times the growth is driven by the source term.

\begin{figure}
\vskip -0.2in \hskip -0.4in
\begin{center}
\includegraphics[scale=0.3,angle=90,width=3.in,clip]{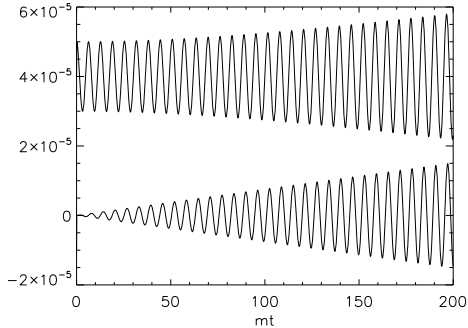}
\end{center}
\vskip -.2in \caption{ Plotted are $\delta \phi$ (upper curve) and $\delta \phi_c$ for the
toy model described in the text with $\phi(0) = 10^{-1}$ and $\delta \phi(0) = 10^{-5}$.
The upper curve has been offset by $4 \times 10^{-5}$.
 }
\label{fig-toy}
\end{figure}

As an illustration it is interesting to first look at a toy model in which $\la \psi^2 \ra$ in~\eqref{sbe} is replaced by the last term in~\eqref{psi2-def} with $\epsilon/M$
chosen so that this term is equal to $-\phi^2/g^2$ and the resulting equation for $\phi$ is $\ddot{\phi} + (1 - \phi^2) \phi = 0$.
Then $g^2 \delta \la \psi^2 \ra = -2 \phi^2$ and the source term for $\delta \phi_c$ is $2 c \phi^3$.  The solutions for $\phi$
are stable for the starting values $0 < \phi(0) < 1$ and $\dot{\phi}(0) = 0$.  In this case it is easy to solve the linear response equation directly. 
The results for $\phi(0) = 10^{-1}$ and $\delta \phi(0) = 10^{-5}$ are shown in Fig.~\ref{fig-toy}.
One sees that over the range shown there is linear growth in the amplitude of $\delta \phi_c$ while the amplitude of $\delta \phi$ does not grow
significantly initially.  This pattern of early growth of $\delta \phi_c$ is also seen in the solutions to the full set of backreaction equations.

\begin{figure}
\vskip -0.2in \hskip -0.4in
\begin{center}
\includegraphics[scale=0.3,angle=90,width=3.in,clip]{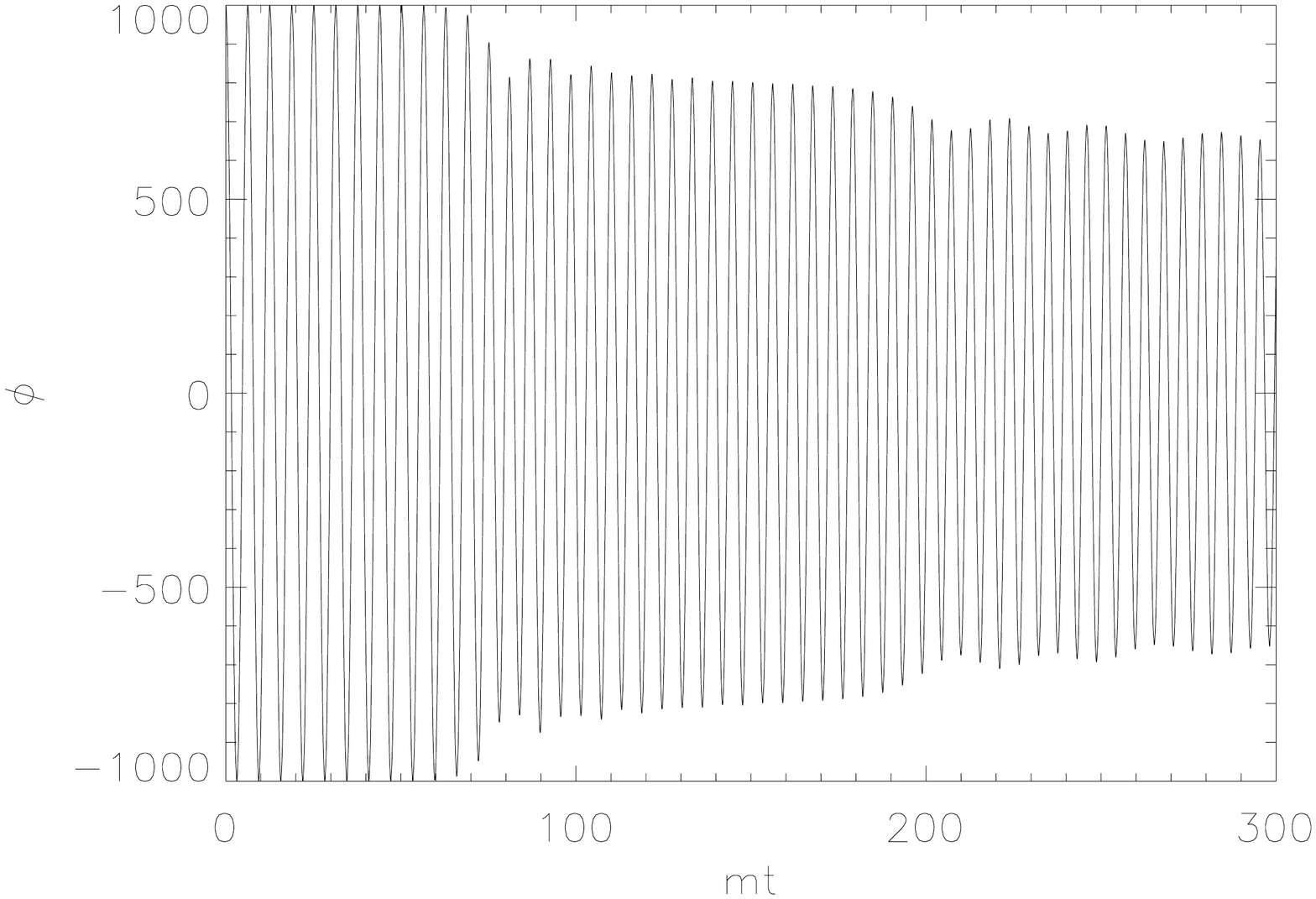}
\includegraphics[scale=0.3,angle=90,width=3.in,clip]{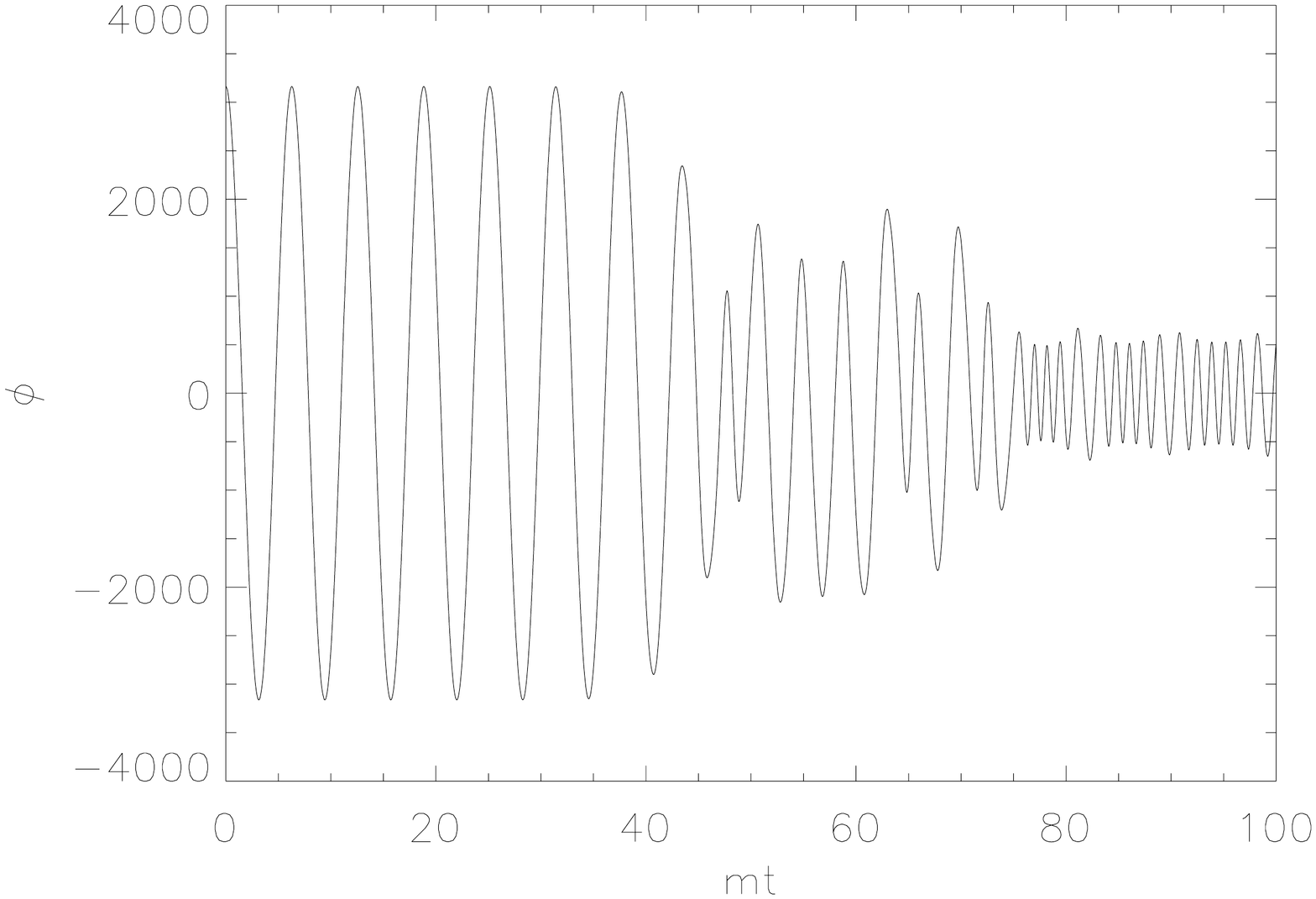}
\includegraphics[scale=0.3,angle=90,width=3.in,clip]{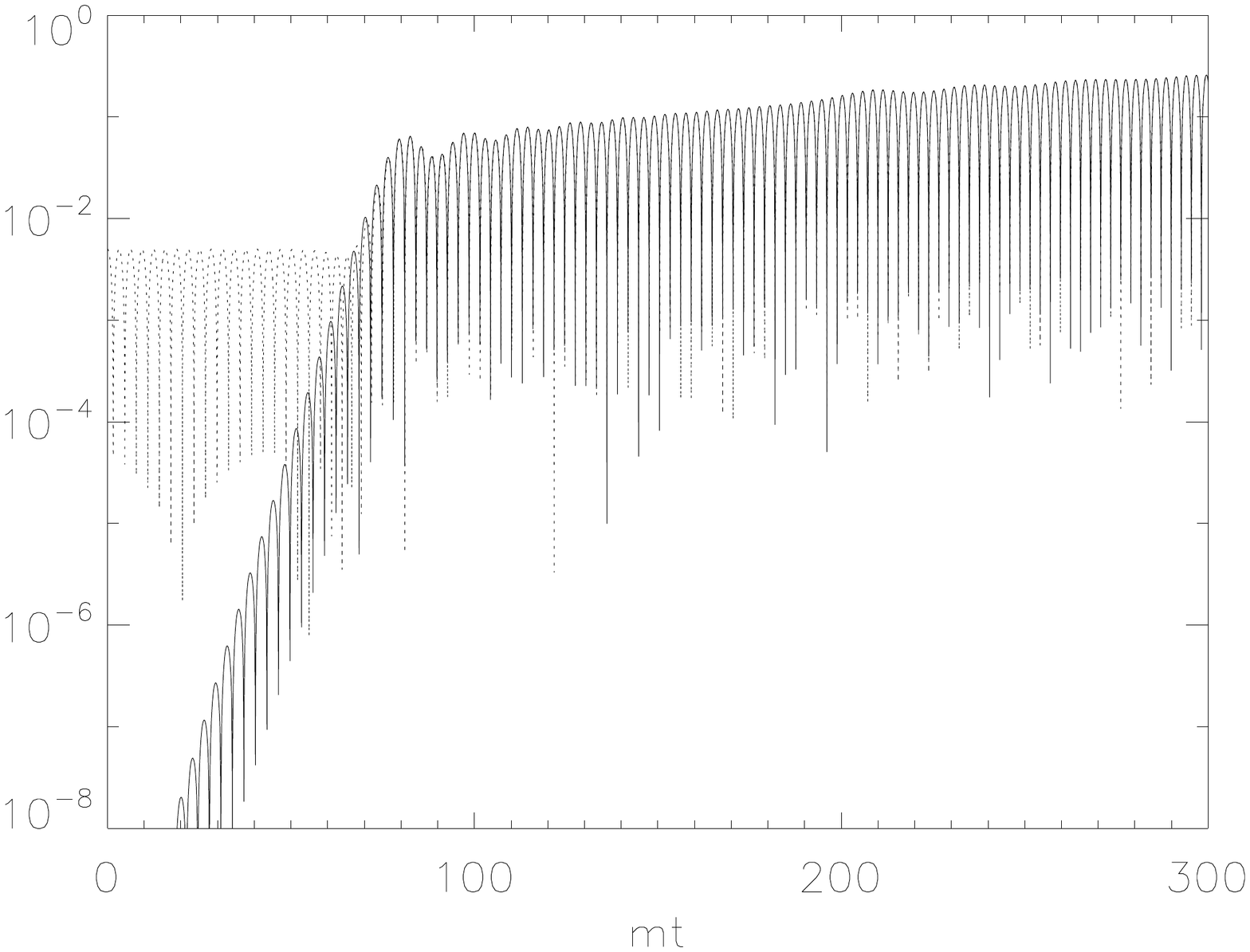}
\includegraphics[scale=0.3,angle=90,width=3.in,clip]{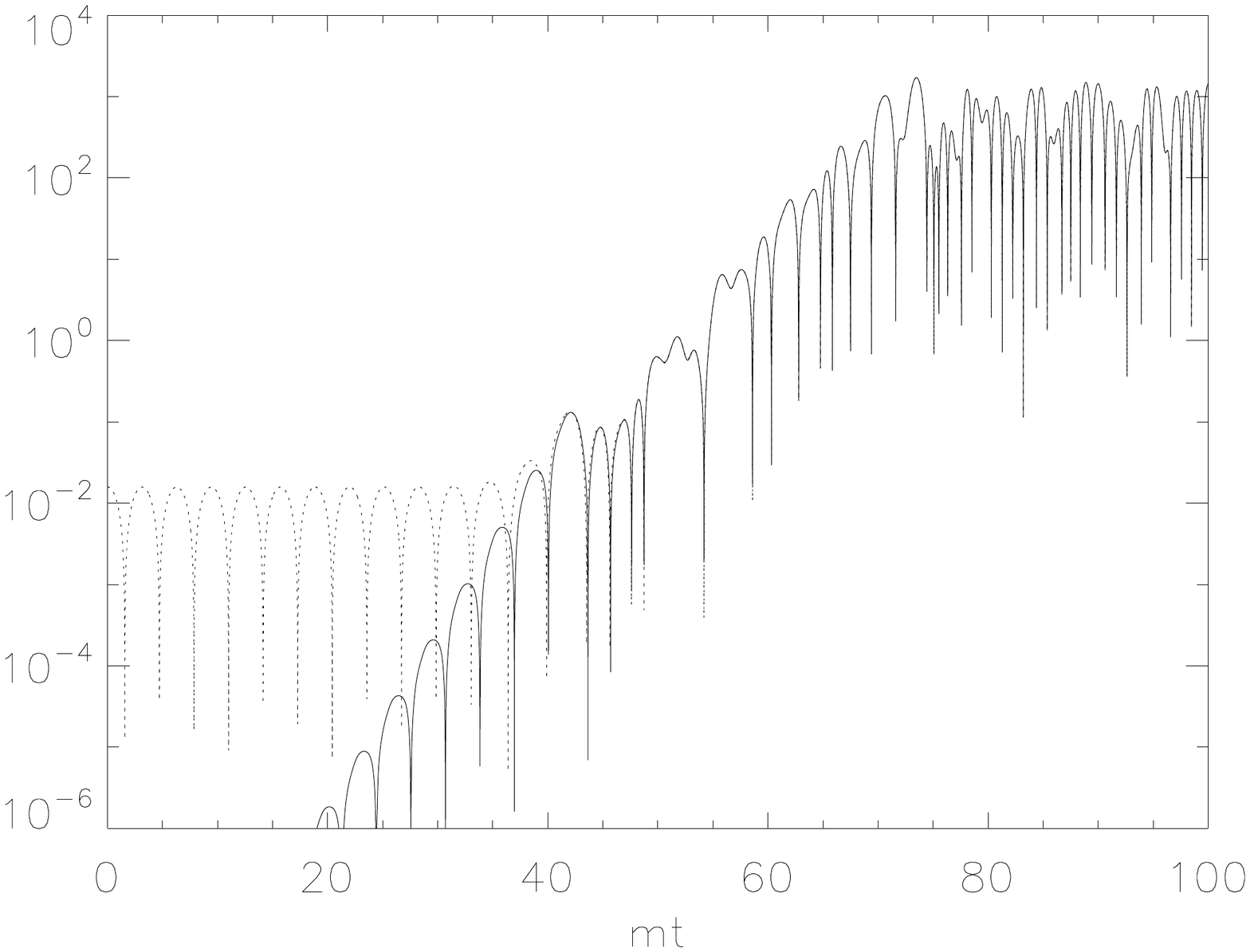}
\end{center}
\vskip -.2in \caption{ The inflaton field $ \phi$ is plotted versus time in the upper panels, and $|\delta \phi_e|$ (dashed curves) and $|\delta \phi_c|$ (solid curves) are plotted versus time in the
lower panels.  For each plot $g = 10^{-3}$.  For the upper left panel $g^2 \phi_0^2 = 1$ and the upper right one $g^2 \phi_0^2 = 10$.
In the lower left panel $\delta \phi_e$ is the difference between solutions with $g^2 \phi_0^2 = 1+10^{-5}$ and $g^2 \phi_0^2 = 1$.  In the
lower right panel $\delta \phi_e$ is the difference between solutions with $g^2 \phi_0^2 = 10(1+10^{-5})$ and $g^2 \phi_0^2 = 10$.
 }
\label{fig-phi}
\end{figure}

In~\cite{kls2} it was predicted that there are two qualitatively different types of solutions to the backreaction equation for $\phi$.
For one there is a relatively slow damping of the inflation field while for the other there is a period in which the inflaton field
is rapidly damped.  In~\cite{amec} it was found for a flat space background that rapid
damping of the inflaton field occurs whenever $g^2 \phi_0^2 \stackrel{_>}{_\sim} 2$ for models in which the starting values are $\phi_0 = \phi(t=0)$
 and $\dot{\phi}(t=0) = 0$.  Rapid damping does not occur for significantly smaller values such
as $g^2 \phi_0^2 = 1$. Whenever rapid damping does occur it is observed to happen twice and there appears to be no significant damping after that.
These effects are illustrated in the upper panels of Fig.~\ref{fig-phi} where the inflaton field is plotted as a function of time for $g^2 \phi_0^2 = 1$ and $g^2 \phi_0^2 = 10$.

Examination of the plots in Fig.~\ref{fig-phi} shows that
in both cases $\delta \phi_e$ grows exponentially at about the time that a significant amount of damping of $\phi$ first occurs, while
$\delta \phi_c$ grows exponentially starting at much earlier times.  After $\delta \phi_c$ grows to be comparable in size
to $\delta \phi_e$ the two quantities are nearly identical and cannot be distinguished on the scale of the plots.
For $g^2 \phi_0^2 = 1$ a small amount of damping of $\phi$ occurs very quickly followed by a much slower damping rate which goes on for a long time. During this latter period $\delta \phi$ grows approximately linearly in time.  For $g^2 \phi_0^2 = 10$ the exponential growth of $\delta \phi_e$ continues through the end of the second rapid damping period
and then all growth appears to cease.

Using the detailed analysis of the particle production in~\cite{amec} we find that the rate of growth of $\delta \phi_c$ appears to
be closely tied to the overall particle production rate.  It is exponential when the particle production rate is high, approximately
linear in time during periods of slow damping when the rate is much smaller, and is negligible after the second rapid damping phase
 when the particle production rate is negligible (in cases where rapid damping occurs).

In the case of preheating the growth rate of the solutions to the linear response equation varies significantly over time.
Therefore the criterion in~\cite{amm-validity} should be modified so that the general form of the criterion is:
in cases where a large $N$ semiclassical approximation is used, the approximation will break down if any linearized gauge invariant quantity
 constructed from solutions to the linear response equations with finite non-singular initial data grows significantly for some period of time.
For the model of preheating considered here, there are no gauge fields, so one can just consider solutions to the linear response equation.

The quantity $\delta \phi_c$ in~\eqref{phic-def} may not be useful in all cases but it is useful in
preheating where we have seen that it is a more sensitive measure of quantum fluctuations than $\delta \phi$ is.  A second modification of the criterion specifically for preheating is that the semiclassical
approximation breaks down if $\delta \phi_c$ grows rapidly for some period of time.  It is clear from our results that this criterion is violated during the early stages of
preheating, well before either scattering effects or backreaction effects
are important.

As pointed out in~\cite{kls2}, the flat space approximation does not always give an accurate account of the details of the preheating process because the expansion
of the universe can have a significant effect on the parametric amplification process.  Nevertheless our results strongly suggest that whenever there is a period
in which a lot of parametric amplification occurs, the semiclassical approximation breaks down.

 This is the first application that has been made of the criterion in~\cite{amm-validity} for the validity of the semiclassical approximation when particle production effects
 are significant.  We think it likely, but cannot be certain, that our results generalize to similar situations and thus that the semiclassical approximation may never
 be valid when there is a high rate of particle production.

\begin{acknowledgments}

P.R.A. and C.M-P. would like to thank Emil Mottola for helpful
conversations.  This work was supported in part by the National
Science Foundation under Grant Nos. PHY-0556292, PHY-0856050, and PHY-1308325 to Wake Forest
University.  The numerical computations herein were performed on the
WFU DEAC cluster; we thank the WFU Provost's Office and Information
Systems Department for their generous support.

\end{acknowledgments}

\end{document}